\documentclass[aip,rsi,reprint,graphicx]{revtex4-1} 

\usepackage{amsfonts,amsmath,mathrsfs} 
\usepackage{dcolumn}
\usepackage{graphicx}
\usepackage{amssymb}
\usepackage{multirow}
\usepackage{setspace}
\usepackage{needspace}

\begin{document}

\title{Ultra-wide frequency response measurement of an optical system with a DC photo-detector} 

\author{Katanya B. Kuntz}
\email[]{katanyab@gmail.com}
\affiliation{School of Engineering and Information Technology, The University of New South Wales, Canberra, ACT 2600, Australia}
\affiliation{Centre for Quantum Computation and Communication Technology, Australian Research Council, Australia}
\affiliation{Institute for Quantum Computing, University of Waterloo, Waterloo, ON, Canada N2L 3G1}
\affiliation{Department of Physics and Astronomy, University of Waterloo, Waterloo, ON, Canada N2L 3G1}

\author{Trevor A. Wheatley}
\affiliation{School of Engineering and Information Technology, The University of New South Wales, Canberra, ACT 2600, Australia}
\affiliation{Centre for Quantum Computation and Communication Technology, Australian Research Council, Australia}

\author{Hongbin Song}
\affiliation{School of Engineering and Information Technology, The University of New South Wales, Canberra, ACT 2600, Australia}
\affiliation{Centre for Quantum Computation and Communication Technology, Australian Research Council, Australia}
\affiliation{School of Humanity and Social Science, The Chinese University of Hong Kong, Shen Zhen, Guang Dong, China}

\author{James G. Webb}
\affiliation{School of Engineering and Information Technology, The University of New South Wales, Canberra, ACT 2600, Australia}
\affiliation{EOS Space Systems Pty Ltd, EOS House, Mt Stromlo Observatory, Cotter Road, Weston
Creek, ACT 2611, Australia}

\author{Mohamed A. Mabrok}
\affiliation{School of Engineering and Information Technology, The University of New South Wales, Canberra, ACT 2600, Australia}
\affiliation{Robotics, Intelligent Systems and Control Lab, Computer, Electrical and Mathematical Science and Engineering, King Abdullah University of Science and Technology (KAUST), Saudi Arabi}

\author{Elanor H. Huntington}
\affiliation{School of Engineering and Information Technology, The University of New South Wales, Canberra, ACT 2600, Australia}
\affiliation{Centre for Quantum Computation and Communication Technology, Australian Research Council, Australia}
\affiliation{Research School of Engineering, College of Engineering and Computer Science, Australian National University, Canberra, ACT 2600, Australia}

\author{Hidehiro Yonezawa}
\affiliation{School of Engineering and Information Technology, The University of New South Wales, Canberra, ACT 2600, Australia}
\affiliation{Centre for Quantum Computation and Communication Technology, Australian Research Council, Australia}
\email[]{h.yonezawa@unsw.edu.au}


\begin{abstract}
Precise knowledge of an optical device's frequency response is crucial for it to be useful in most applications. Traditional methods for determining the frequency response of an optical system (e.g. optical cavity or waveguide modulator) usually rely on calibrated broadband photo-detectors or complicated RF mixdown operations. As the bandwidths of these devices continue to increase, there is a growing need for a characterization method that does not have bandwidth limitations, or require a previously calibrated device. We demonstrate a new calibration technique on an optical system (consisting of an optical cavity and a high-speed waveguide modulator) that is free from limitations imposed by detector bandwidth, and does not require a calibrated photo-detector or modulator. We use a low-frequency (DC) photo-detector to monitor the cavity's optical response as a function of modulation frequency, which is also used to determine the modulator's frequency response. Knowledge of the frequency-dependent modulation depth allows us to more precisely determine the cavity's characteristics (free spectral range and linewidth). The precision and repeatability of our technique is demonstrated by measuring the different resonant frequencies of orthogonal polarization cavity modes caused by the presence of a non-linear crystal. Once the modulator has been characterized using this simple method, the frequency response of any passive optical element can be determined.
\end{abstract}
     
\maketitle 


\section{Introduction}

It is imperative to know the frequency response of an optical device (e.g. a cavity or high-speed modulator) for it to be useful in optics or communication applications. For example, the frequency spacing between resonances (the free spectral range, FSR) of a cavity must be known before being used to calibrate a laser's frequency \cite{Adam1989, Gamache1996}, perform absolute length measurements \cite{Bay1968,Haitjema2000,Lawall2005}, generate non-classical light \cite{Senior2007}, or produce large cluster states using frequency-entangled photons \cite{Pysher2011,Chen2014,MedeirosdeAraujo2014}. Dense wavelength division multiplexed systems often use Fabry-Perot etalons to select and stabilize the wavelength of a tunable diode laser \cite{Williamson2004,Gee2006}. A precise measurement of the etalon's FSR is necessary for matching its transmission channel with the International Telecommunication Union grid. Similarly, it is necessary to calibrate a high-speed amplitude modulator before use in long-haul optical fiber transmission systems \cite{Korotky1987b,Okiyama1988,Wooten2000}, or as a wideband optical signal source for characterizing the frequency response of photo-detectors or optical fibers \cite{Kobayashi1976,Ito1976}. 

Several methods exist for measuring the frequency response of an optical cavity or modulator. The most-straightforward method is direct detection with a calibrated fast receiver. However, this technique is highly dependent on the bandwidth of the optical receiver, and the ability to calibrate it. Most characterization techniques involve a combination of either a calibrated wide bandwidth optical receiver, a commercial optical spectrum analyzer or a calibrated high-speed modulator. For example, techniques based on frequency modulation are effective in determining a cavity's FSR \cite{Uehara1995,Manson1999,Ozdur2008,Mandridis2010,Aketagawa2011}. However, these methods require both a calibrated modulator and a fast photo-detector with a known frequency response. Several techniques exist that can determine the response of a modulator that do not require a fast photo-detector, such as optical heterodyning \cite{Tan1989,Hawkins1991,Lam2006,Chtcherbakov2007} or swept-frequency techniques \cite{Uehara1978}. Unfortunately, these methods still rely on fast electronics or specialized laser sources that only operate at particular wavelengths. As optical device bandwidths continue to increase, it is necessary to develop a measurement technique that determines their frequency response without depending on the precise calibration of a high-speed photo-detector, modulator, or optical cavity. 

In this paper, we use a recursive method to obtain the frequency response of an optical system. This system consists of both an optical cavity and a high-speed (uncalibrated) amplitude modulator, which can be simultaneously characterized from the same measurement configuration. Our approach extends a technique demonstrated by Locke \textit{et al.}  \cite{Locke2009}, and involves a fiber-coupled amplitude (intensity) modulator, and a low-frequency (DC) photo-detector, both of which are commercial laboratory equipment. The photo-detector is used to measure the DC cavity transmitted or reflected power as a function of the modulation frequency, while the carrier remains frequency locked to a resonance. Instead of characterizing different cavity resonances by shifting the center laser wavelength \cite{Locke2009}, we characterize the full bandwidth up to 15.5$\:$GHz from the optical carrier (28 resonances) by locking the center wavelength, and only adjusting the modulation frequency. By measuring the entire cavity response (both on and off-resonance) over this frequency range, we are able to estimate the frequency-dependent modulation depth, and use this knowledge to more precisely determine the cavity characteristics (FSR and linewidth). In addition, we explore the precision of our technique by quantifying the effect modulation harmonics have on the measured linewidth, and determining the different resonant frequencies of orthogonal polarization cavity modes caused by the presence of a birefringent material (non-linear crystal). Once the modulator has been calibrated, this measurement technique can be used to characterize the frequency response of any passive optical element, and is not limited to cavities. 

\section{Theory}
\label{sec:theory}

In the next three subsections, we will present our proposed method, theories related to characterizing optical cavities, and how to estimate the frequency response of an optical cavity and amplitude modulator.

\subsection{Proposed method}

\begin{figure*}
\centering{\includegraphics[height=5 cm]{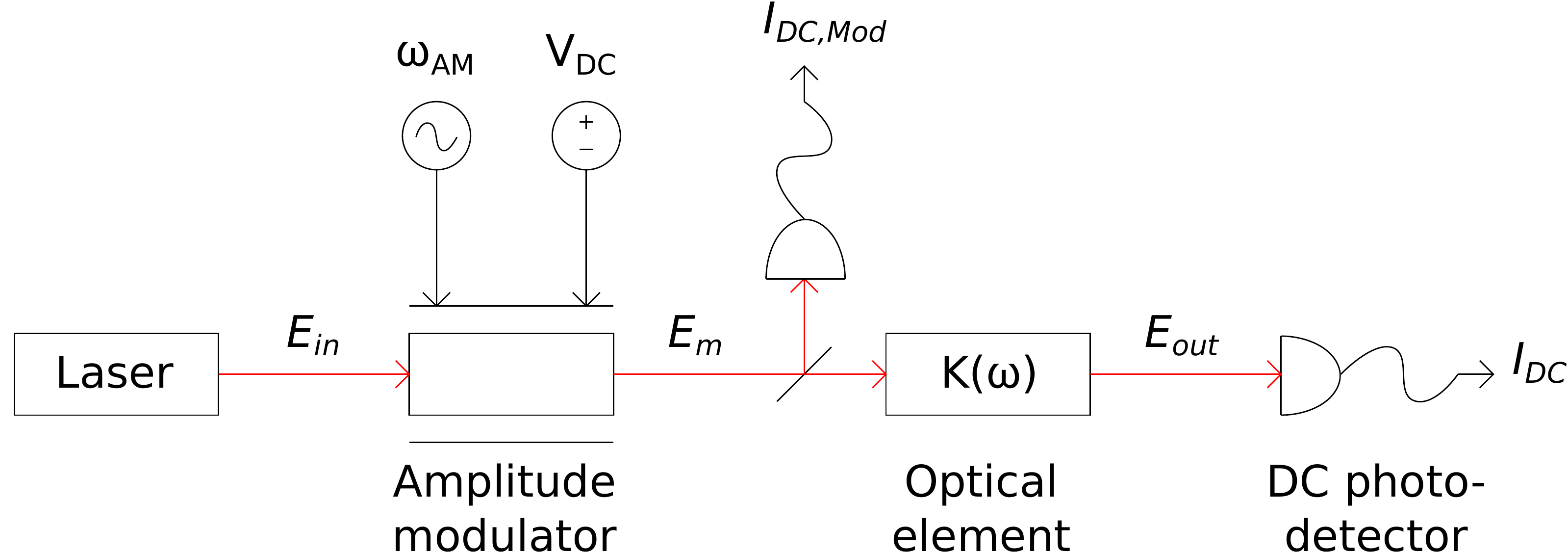}}
\caption{Schematic representing the model of our proposed method to measure the frequency response of an optical system consisting of an optical cavity and an amplitude modulator. First a laser with electric field amplitude $E_{in}$ is sent to the modulator. The amplitude modulated light ($E_m$) is then injected into the optical element under investigation, whose transmission/reflection coefficient is given as $K(\omega)$. The light from the optical element ($E_{out}$) is detected with a low-frequency (DC) photo-detector. The detector output is described by $I_{DC}$ in Eq. \ref{eqn:Idc}. A small pick-off of the modulator's output is also monitored with a DC photo-detector, whose output is described by $I_{DC,Mod}$ in Eq. \ref{eqn:Idc0}. The modulator is driven with a sine wave signal at $\omega_{AM}$, and biased with a DC voltage supply at $V_{DC}$.} \label{fig:Theorysetup}
\end{figure*}

In this section, we will briefly explain our proposed method, which is shown in Fig. \ref{fig:Theorysetup}. The setup consists of a laser, a high-frequency amplitude modulator, an optical element to be measured, and low-frequency (DC) photo-detectors. We first use a cavity as the optical element under investigation as we can use the cavity's frequency response to calibrate the modulator's response. Once the modulator has been calibrated, our method can be applied to characterize any optical element to a fine resolution (e.g., kilohertz) over tens of gigahertz. 

First we modulate the laser beam with an amplitude modulator. The electric field of the output of the amplitude modulator, $E_m(t)$, can be written as
   \begin{align}
      E_m(t)=\frac{1}{2}E_{in}e^{i\omega_0 t}  (1+e^{i \left[ \theta_{DC} +\theta_{AC}(t) \right]}),
    \label{eq:Em}
   \end{align}
where $E_{in}$ is the amplitude of the input to the amplitude modulator, $\omega_0$ is the laser carrier angular frequency, $\theta_{DC}$ and $\theta_{AC}$ are phase shifts caused by DC and AC voltages applied to the amplitude modulator, respectively. Note that here we assume that the amplitude modulator consists of a Mach Zehnder interferometer with a phase modulator in one of two paths, but our derivation will also be valid for other amplitude modulator configurations \cite{Saleh2007}.

The AC signal applied to the modulator is a sine wave with an angular frequency $\omega_{AM}$, as shown in Fig. \ref{fig:Theorysetup}. The phase shift caused by this AC signal is given as $\theta_{AC} = \beta \sin(\omega_{AM}t)$, where $\beta$ is the modulation depth. The modulation depth is usually frequency dependent, as we will later show is the case in our experiment. The electric field $E_m(t)$ is now written as
\begin{align} \nonumber
E_m(t)=&\: \frac{1}{2} E_{in}e^{i\omega_0 t} 
              ( 1 + e^{i\theta_{DC}} e^{i \beta \sin(\omega_{AM}t) })  \\
=&\: \frac{1}{2} E_{in}e^{i\omega_0 t} 
              \left( 1 + e^{i\theta_{DC}}\sum_{n=-\infty}^{+\infty} J_n(\beta) e^{i n \omega_{AM}t}
              \right),
    \label{eq:Em2}
   \end{align}
where $J_n(\beta)$ are $n^{\mathrm{th}}$-order Bessel functions of the first kind. 

The modulated beam is then injected into the optical element under investigation, and the output (either transmission or reflection) is measured with a low-frequency (DC) photo-detector, as depicted in Fig. \ref{fig:Theorysetup}. In order to calculate the output signal, we apply a Fourier transform to the modulator output $E_m(t)$, which gives
\begin{align} \nonumber 
E_m(\omega)=&\: \frac{1}{2} E_{in} 
              \bigg( 2 \pi \delta(\omega-\omega_0) \\
+&\: e^{i\theta_{DC}} \sum_{n=-\infty}^{+\infty} J_n(\beta)
               2 \pi \delta(\omega- \omega_0- n \omega_{AM} \bigg).
    \label{eq:Em2w}
\end{align}
After passing through the optical element, whose transmission/reflection coefficient (transfer function) is given as $K(\omega)$, the output electric field is written as 
\begin{align} \nonumber
E_{out}(\omega) =&\: E_m(\omega)K(\omega), \\ \nonumber
=&\:\frac{1}{2} E_{in} \bigg(2 \pi \delta(\omega-\omega_0) K(\omega) \\
+&\: e^{i\theta_{DC}} \sum_{n=-\infty}^{+\infty} J_n(\beta)
                2 \pi \delta(\omega- \omega_0 - n \omega_{AM}  ) K(\omega)\bigg).
    \label{eq:Eout2}
\end{align}
Note that we ignored the pick-off beam splitter in Fig. \ref{fig:Theorysetup} as it only changes the power scaling of the detected signal. By applying the inverse Fourier transform, we obtain
\begin{align} \nonumber
E_{out}(t)=&\: \frac{1}{2} E_{in}  e^{i\omega_0 t} 
              \bigg( K(\omega_0) \\
+&\: e^{i\theta_{DC}} \sum_{n=-\infty}^{+\infty} J_n(\beta)
                K( \omega_0 + n \omega_{AM}) e^{i n \omega_{AM} t} \bigg).
    \label{eq:Eout}
   \end{align}
The output electric field is measured by a DC photo-detector whose cut-off frequency is much lower than the modulation frequency, $\omega_{AM}$. The detector output is given by
\begin{align} \nonumber 
I_{DC}=&\:G_{det} |E_{out}(t)E_{out}^*(t)|^2, \\ \nonumber
\simeq&\: \frac{I_0}{2}\bigg(|K(0)|^2\big(1+|J_0(\beta)|^2+2J_0(\beta)\:\mathrm{cos}\:\theta_{DC}\big) \\
 +&\: \sum_{n\neq 0} |J_n(\beta)|^2|K(n\omega)|^2\bigg),
     \label{eqn:Idc}
   \end{align}
where $G_{det}$ is the DC gain of the detector, $I_0 = \frac{G_{det} |E_{in}|^2}{2}$, and we let $\omega_0=0$ and $\omega_{AM} \rightarrow \omega$. This is the general expression of our method, where the modulation depth, $\beta$, and the coefficient $K(\omega)$ are frequency dependent. We also note that the laser power ($I_0$) and the DC offset ($\theta_{DC}$) may fluctuate during experiments. In particular, internal heating due to conductive losses in the modulator may change the DC offset. Since this loss is frequency dependent, the DC offset can drift with changing modulation frequency. Our procedure makes it possible to precisely determine both $\beta$ and $K(\omega)$ in the presence of these fluctuations. Before detailing our procedure, we will briefly present a mathematical description of an optical cavity in the next section. 

\subsection{Measurements of optical cavities}

In this section, we will briefly explain the theories related to the characterization of optical cavities. Firstly, the transmission $K_T(\omega)$ and reflection $K_R(\omega)$ of ideal cavities are written as \cite{Saleh2007}
    \begin{align} \label{eq:TandR}
       K_T(\omega) &= \frac{ 
       					   e^{i\phi/2}	\sqrt{ (1-L')(1-R_{in})(1- R_{out}) }
       						       					} 
                        { 
                        	1 - e^{i\phi} \sqrt{ ( 1-L ) R_{in} R_{out}  }  
                        }, 
       \\
        K_R(\omega) &= \frac{
               					-\sqrt{R_{in} } + e^{i\phi} \sqrt{ (1-L) R_{out} } 
         			 	  } 
             			 { 
             				 1 - e^{i\phi} \sqrt{ (1-L) R_{in} R_{out} }  
              			}, 
       \label{eq:TandR2}
	\end{align}
where $R_{in}$ and $R_{out}$ are intensity reflection coefficients for the input and output mirrors, respectively, and $L$ is the intra-cavity loss for a round trip ($L'$ is the loss for a half trip, $L'\sim L/2$ ). $\phi$ is the phase shift for a round trip of the cavity, which is given as $\phi=(\omega_0+\omega)l/c$, where $c$ is a speed of light, and $l$ is the effective length of the cavity.
 
We assume the laser carrier is locked to the cavity (i.e., $e^{i \omega_0 l /c} = 1$). The cavity is on resonance when $\omega = n \omega_{FSR}$, where  $ \omega_{FSR}$ is the resonant angular frequency, which satisfies $e^{i n \omega_{FSR} l / c } = 1$, and $n$ is an integer. Near a cavity resonance (i.e., $\omega \approx n\omega_{FSR}$ ), $|K_T(\omega)|^2$ and $|K_R(\omega)|^2$ can be approximated to a Lorentzian function as
    \begin{align} \label{eq:T2}
    |K_T(\omega)|^2 \simeq &\: \frac{\gamma^2} {(\omega-n\omega_{FSR})^2+\gamma^2} 
                      \times  |K_T(n\omega_{FSR})|^2,
    \\ \nonumber 
    |K_R(\omega)|^2 \simeq &\: 1 - \bigg( \frac{\gamma^2}{(\omega-n\omega_{FSR})^2+\gamma^2} \\ \label{eq:R2}
                      \times&\:  \left( 1-|K_R(n\omega_{FSR})|^2 \right ) \bigg),        
    \\ \nonumber
    \gamma =&\: \pi f_{FWHM} \\
    =&\: \frac{1- \sqrt{R_{in}R_{out}} \sqrt{1 - L}  }
                                       { [ R_{in}R_{out} ( 1 - L)]^{1/4} } 
                                       f_{FSR},
    \\
    f_{FSR} =&\: \frac{c}{l},
 	\end{align}
where $|K_T(n\omega_{FSR})|^2$ and $|K_R(n\omega_{FSR})|^2$ are the (constant) intensity transmission and reflection coefficients at the resonance, respectively, $f_{FSR}$ is the free spectral range (FSR), and $f_{FWHM}$ is the linewidth (full width at half maximum, FWHM). 

\begin{figure}
\centering{\includegraphics[width=9 cm]{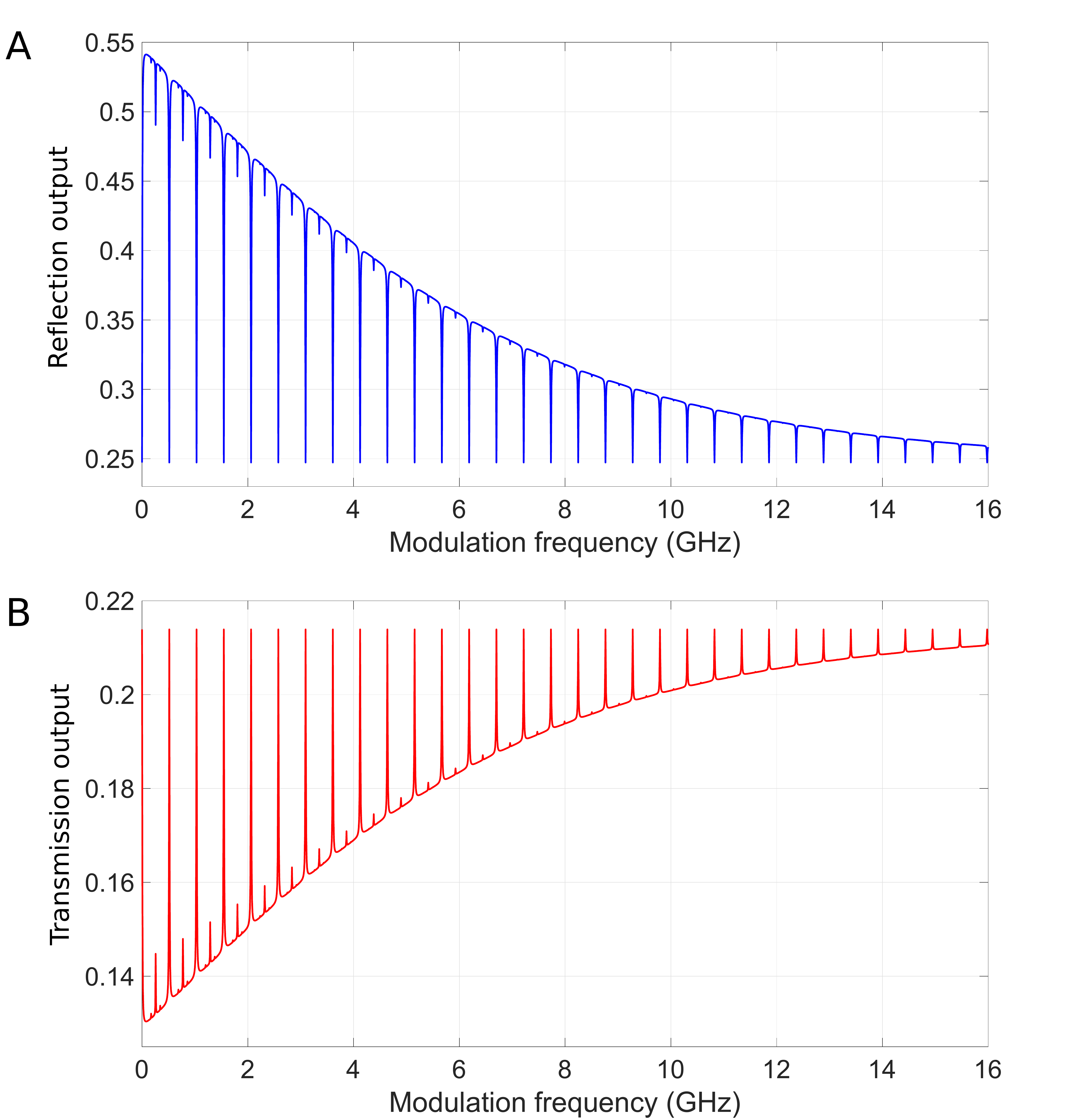}}
\caption{Theoretical calculations for the cavity reflection (A) and transmission (B) measurements. The overall downward (upward) trend of the off-resonance data in the reflected (transmitted) response, located inbetween the resonances, is due to the increasing optical power at the carrier frequency as a function of the modulator's frequency response. We used parameters similar to the characteristics of our optical cavity. We modelled the frequency-dependent modulation depth of the amplitude modulator as $\beta=1.5 e^{-f/8.69\:\mathrm{GHz}}$ ($-20\:$dB at $20\:$GHz), normalized the incident intensity as $I_0=1$, and set the DC offset of the modulator to $\theta_{DC} = \pi/2$.} \label{fig:Theoryplot}
\end{figure}

Figure \ref{fig:Theoryplot} shows the theoretical detector outputs for a cavity under investigation using our method, which is derived by substituting Eq. \ref{eq:TandR} (or Eq. \ref{eq:TandR2}) into Eq. \ref{eqn:Idc}. In this case, the transmission $K_T(\omega)$ and reflection $K_R(\omega)$ are symmetric, that is, $K_T(\omega)=K_T^*(-\omega)$ and $K_R(\omega)=K_R^*(-\omega)$. We chose similar model parameters to our experimental values. The intensity reflection coefficients for the input and output mirrors are $R_{in}=0.9$ and $R_{out}=0.99$, respectively. The total intra-cavity loss is $L=0.025$, and the effective path length is $l=0.582\:$m. We also assume that the modulation depth, $\beta$, is frequency dependent as $\beta=\beta_0 e^{-f/f_0}$, and that the laser power and DC offset of the modulator are constant for simplicity. The input optical power to the amplitude modulator is distributed between the power at the carrier frequency and at the sideband frequencies. If the input power is constant during the measurement, then the sum of the optical powers at these frequencies will also be constant. As a result of the modulator's frequency response, if the modulation signal strength is held constant while the frequency is swept, the optical power in the sidebands will decrease for higher modulation frequencies. Thus the optical power at the carrier frequency will increase. The overall upward (downward) trend of the off-resonance data in the transmitted (reflected) data, located inbetween the resonances, is due to the increasing power at the carrier frequency as a function of the modulator's frequency response. Also, evidence of the second harmonic modulation sidebands from the amplitude modulator coupling into the cavity is visible as smaller peaks located halfway between the main resonances. In the next section, we will present the mathematical description of our measurement procedure.

\subsection{Frequency response measurement of an optical cavity and amplitude modulator}

In this section, we will derive how to estimate the frequency response of an optical cavity and the frequency-dependent modulation depth, $\beta$, of a high-speed amplitude modulator. First we scan the modulation frequency over tens of gigahertz to determine the responses of the cavity and modulator. During the frequency scan, the laser power may fluctuate, as well as the DC offset of the waveguide modulator may change due to internal mechanisms (e.g. internal heat effects via pyroelectric effect \cite{Skeath1986}, or changing environmental conditions \cite{Jungerman1990}). We assume that the DC offset is set to $\pi/2$ at the start of the scan (i.e., at low frequency), and that the modulator has no significant acoustic resonances over the measurement bandwidth. We also assume that the modulation depth is approximately constant over the cavity resonance (which is $\sim10\:$MHz). Under these assumptions, we can first estimate the resonant frequencies and linewidth of cavity from the data, and then determine the modulator response by using the cavity as a reference. Finally, we determine the cavity characteristics more precisely using the information about the modulator response. 

First we derive an expression for the modulator's output, which is monitored using a DC photo-detector, as depicted in Fig. \ref{fig:Theorysetup}. The output of the modulator is derived by letting $|K(\omega)| = 1$ in Eq. \ref{eqn:Idc}, which gives
\begin{align}
I_{DC,Mod} \:=&\:I^\prime_0\big(1+J_0(\beta)\:\mathrm{cos}\:\theta_{DC}\big), \label{eqn:Idc0}
\end{align}
where $I^\prime_0$ is proportional to the incident laser power, $I_0$ (i.e. $I^\prime_0 \propto I_0$).

Next we will look at the transmission or reflection of a cavity. In order to eliminate the effects from laser power fluctuations, we normalize Eq. \ref{eqn:Idc} with Eq. \ref{eqn:Idc0}, and define $\mathscr{K}= I_0/I^\prime_0$, which is constant. The normalized DC detector output becomes
\begin{align} \nonumber
\mathscr{I}_{DC} \:=&\: \frac{\mathscr{K}}{2} \frac{1}{1+J_0(\beta)\:\mathrm{cos}\:\theta_{DC}} \\ \nonumber
\times& \bigg(|K(0)|^2\big(1+|J_0(\beta)|^2+2J_0(\beta)\:\mathrm{cos}\:\theta_{DC}\big) \\
+&\: \sum_{n\neq 0} |J_n(\beta)|^2|K(n\omega)|^2\bigg).
\label{eqn:IdcNorm}
\end{align}

Once we measure $\mathscr{I}_{DC}$ over a wide frequency range, we can determine the FSR of the cavity from the resonance peaks visible in Fig. \ref{fig:Theoryplot}. To determine the linewidth, we may assume that $\beta$ and $\theta_{DC}$ are approximately constant over the linewidth, which gives
\begin{equation}\label{eqn:IdcPrime}
\mathscr{I}_{DC}^\prime =\: \mathscr{K}^\prime \sum_{n=-\infty}^{+\infty}  |J_n(\beta)|^2 |K(n\omega)|^2 + \mathcal{C},
\end{equation}
where $\mathscr{K}^\prime$ and $\mathcal{C}$ are constants.

We can use the derived expressions of $K_T(\omega)$ and $K_R(\omega)$ to approximately fit the data and obtain the linewidth. If the modulation depth, $\beta$, is unknown at this point, we may use $\beta$ as a fitting parameter, and neglect $|n|\geq2$ to realize a reasonable fit. Next we will extract $\beta$ over the measured frequency range, and then revisit the fitting to improve our precision in determining the FSR and linewidth. 

We can extract the on-resonance and off-resonance data from $\mathscr{I}_{DC}$ to determine the modulation depth. If the modulation sidebands are on resonance with the cavity, then $|K(n\omega)| = K(0)$. The normalized on-resonance detector output is
\begin{align} \nonumber
\mathscr{I}_{DC}^{on} \:=&\: \frac{\mathscr{K}}{2} \frac{1}{1+J_0(\beta)\:\mathrm{cos}\:\theta_{DC}} \\ \nonumber
\times& \bigg(|K(0)|^2\big(1+|J_0(\beta)|^2+2J_0(\beta)\:\mathrm{cos}\:\theta_{DC}\big) \\
+&\: \sum_{n\neq 0} |J_n(\beta)|^2|K(0)|^2\bigg), \nonumber \\
=&\:\mathscr{K} |K(0)|^2. \label{eqn:IdcON}
\end{align}
This expression indicates that the on-resonance detector response is constant regardless of the modulation depth, $\beta$, and the DC offset, $\theta_{DC}$. We can estimate $\mathscr{I}_{DC}^{on}$ from the multiple resonance peaks visible in the cavity's frequency response, as illustrated in Fig. \ref{fig:Theoryplot}. Note that $\mathscr{I}_{DC}^{on}$ is constant over the entire measurement bandwidth.

If the sidebands are off resonance, and we measure the transmitted power, then $|K(n\omega)| \approx0$ and we find,
\begin{equation}
\mathscr{I}_{DC}^{off,T} \:=\: \mathscr{K}|K(0)|^2 \bigg(1+ \frac{|J_0(\beta)|^2 - 1}{2(1+J_0(\beta)\:\mathrm{cos}\:\theta_{DC})}\bigg). \label{eqn:IdcOFF}
\end{equation} 
This off-resonance data can be obtained from $\mathscr{I}_{DC}$ by eliminating data around the resonant frequencies, and then interpolating. By interpolating the data (assuming the modulation depth and the DC offset are approximately constant within the resonance linewidth), we can obtain continuous off-resonance data. Combining Eq. \ref{eqn:IdcON} and \ref{eqn:IdcOFF}, we get
\begin{equation}
\frac{\mathscr{I}_{DC}^{off,T}}{\mathscr{I}_{DC}^{on}}-1 = \frac{|J_0(\beta)|^2-1}{2(1+J_0(\beta)\:\mathrm{cos}\:\theta_{DC})}.
\label{eqn:IdcRatio}
\end{equation}
Since we can assume that $\theta_{DC} = \pi/2$ at $\omega = 0$, the DC detector output of the modulator's response at $\omega = 0$ is $I_{DC,Mod|\omega=0} = I^\prime_0$. Therefore, the normalized DC detector output of the modulator's response without the presence of a cavity is
\begin{equation}
\mathscr{I}_{DC,Mod} = \frac{I_{DC,Mod}}{I_{DC,Mod|\omega=0}} = 1 + J_0(\beta)\:\mathrm{cos}\:\theta_{DC}.
\label{eqn:IdcMod}
\end{equation}
Combining Eq. \ref{eqn:IdcRatio} and \ref{eqn:IdcMod}, we obtain
\begin{equation}
|J_0(\beta)|^2 = 2\mathscr{I}_{DC,Mod} \bigg(\frac{\mathscr{I}_{DC}^{off,T}}{\mathscr{I}_{DC}^{on}}-1\bigg) + 1. \label{eqn:beta}
\end{equation}

\begin{figure*}
\centering{\includegraphics[height=6 cm]{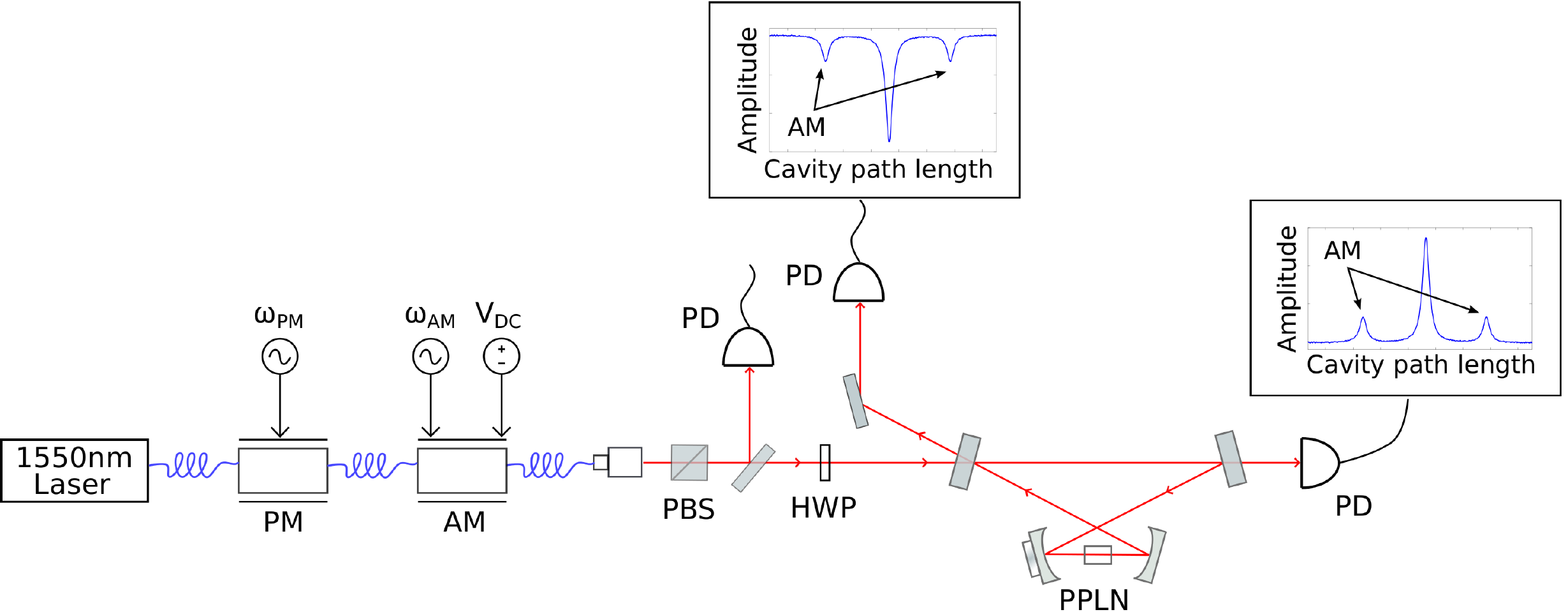}}
\caption{Diagram of our experiment used to measure the frequency response of our optical system consisting of an amplitude modulator and optical cavity with a non-linear crystal (periodically poled lithium niobate, PPLN). Phase modulation (PM) at $\omega_{PM}$ is used to lock the cavity to the laser carrier frequency, while amplitude modulation (AM) at $\omega_{AM}$ is used to probe the cavity's response. A DC bias voltage, $V_{DC}$, is applied to control the energy ratio between the AM sidebands and the carrier. Low-frequency (DC) photo-detectors (PD) are used to monitor the transmitted and reflected power as a function of $\omega_{AM}$. PBS: polarization beam splitter. HWP: half-wave plate.} \label{fig:ExptDiagram}
\end{figure*}

Using this expression, we can determine the optical modulator response, $\beta$, over the entire measurement bandwidth. Once we obtain $\beta$, we can revisit our derivation of the linewidth and FSR frequency to more precisely determine these values. From Eq. \ref{eq:T2}, \ref{eq:R2} and \ref{eqn:IdcPrime}, we can derive an expression for the lineshape of a cavity resonance as
\begin{align} \nonumber
\mathcal{S} \:=&\: J_0(\beta)^2 + 2J_1(\beta)^2 \frac{\gamma^2} {(\omega_{AM}-n\omega_{FSR})^2+\gamma^2} \\
 +&\: 2J_2(\beta)^2 \frac{\gamma^2} {4(\omega_{AM}-n\omega_{FSR})^2+\gamma^2}. \label{eqn:Lineshape}
\end{align}
Note here we use $\omega_{AM}$ rather than $\omega$ to clarify the modulation frequency. This function takes the form of a constant plus two Lorentzian functions. The first function, scaled by $J_1(\beta)$, represents the first-order modulation sidebands being on resonance with the cavity, and has a bandwidth equal to the linewidth, $f_{FWHM}$. While the second function, scaled by $J_2(\beta)$, represents the second harmonic sidebands being on resonance, and has a bandwidth of $f_{FWHM}/2$. Since the modulation depth is not expected to be strong enough to excite the third harmonic, we can ignore modulation sidebands of order $|n|\geq3$.

Therefore, our technique can characterize an optical system that consists of a cavity and a modulator with a single measurement configuration. The on and off resonance data from the cavity measurements is used to learn about the frequency-dependent modulation depth, $\beta$. Then we use $\beta$ to fit the Lorentzian lineshape described by Eq. \ref{eqn:Lineshape} to the on-resonance data to give a more precise FSR and linewidth. A clear advantage of this method is that we only require a DC photo-detector. We do not need an expensive or elaborate high-frequency detector, or to calibrate the frequency response of the detector or modulator beforehand. In the next section, we will present experimental results from applying this technique to characterize an optical system, consisting of a waveguide modulator and an optical cavity with a non-linear crystal.

\section{Measurement setup}

Our frequency response measurement technique involves a recursive approach that allows us to simultaneously characterize both the optical element under investigation (cavity) and the measurement device (amplitude modulator). Thus both the cavity and modulator are characterized from a single setup, and the technique is not restricted by, or dependent on, the frequency response of the photo-detector. We determine the frequency response of our optical system using the experimental set-up shown in Fig. \ref{fig:ExptDiagram}. The output from a 1550nm fiber laser is first sent through a phase modulator, followed by an amplitude modulator. Both devices are fiber-coupled broadband low-loss lithium niobate electro-optical modulators (EOSPACE). The amplitude modulation frequency, $\omega_{AM}$, is varied while the light before and after the cavity is monitored using low-frequency (DC) photo-detectors. The cavity has a birefringent non-linear crystal inside, which has different refractive indices for horizontally and vertically-polarized light. This results in non-degenerate polarization modes that have distinguishable resonant frequencies. Since we use a polarization beam splitter to send linearly-polarized light to the cavity, we can measure either horizontal or vertical polarization modes by simply rotating a half-wave plate before the cavity. The incident power of the laser is quite low (a few milliwatts) to avoid exciting second harmonic generation from the non-linear crystal.

The cavity is locked on resonance with the laser optical carrier frequency during the frequency measurements using a phase modulator and the Pound-Drever-Hall locking technique  \cite{Drever1983}. It must remain locked on resonance while the amplitude modulation frequency is swept to ensure that the cavity is not disturbed by any environmental noise while the data is captured. The cavity's response to the amplitude modulation as a function of cavity path length is illustrated in Fig. \ref{fig:ExptDiagram}. This graph depicts the energy ratio between the sidebands and carrier during our measurements. We can control this energy ratio in two ways: the AC modulation signal strength, and the DC voltage applied to a Mach-Zehnder interferometer inside the amplitude modulator. We operate the modulator near the DC quadrature point ($\theta_{DC} = \pi/2$) to ensure a linear response. As mentioned in the theory section, we found that the DC offset changes during the frequency sweep. Our calibration method allows us to eliminate the effect of the drifting DC offset.

One limitation of this technique is it will not work if both the measurement setup (amplitude modulator) and the optical element (cavity) are completely unknown. We have to assume that the modulator's response is relatively smooth with no acoustic resonances around the cavity resonant frequencies, and that the variation in modulation depth over the linewidth of the cavity is small. Fortunately, the S21 response data provided by the manufacturer with the modulator can be used to determine whether these assumptions are valid (they are valid for our modulator). We also need to assume that the cavity has a Lorentzian-type response with periodic resonances. This can be easily determined by using the DC photo-detectors to monitor the cavity's transmitted or reflected light as a function of cavity path length, as illustrated in Fig. \ref{fig:ExptDiagram}. Once the modulator has been characterize using the cavity's response, this technique can be used to determine the frequency response of any passive optical element including, but not limited to, cavities. We will present the results from our frequency response measurements in the next section.

\section{Results}

In the next three subsections, we will present wide bandwidth measurements of the frequency response of our cavity and amplitude modulator, and how we use this information to more precisely determine the cavity's FSR and linewidth.

\subsection{Wide bandwidth measurement of our optical system}

\begin{figure}
\centering{\includegraphics[height=9 cm]{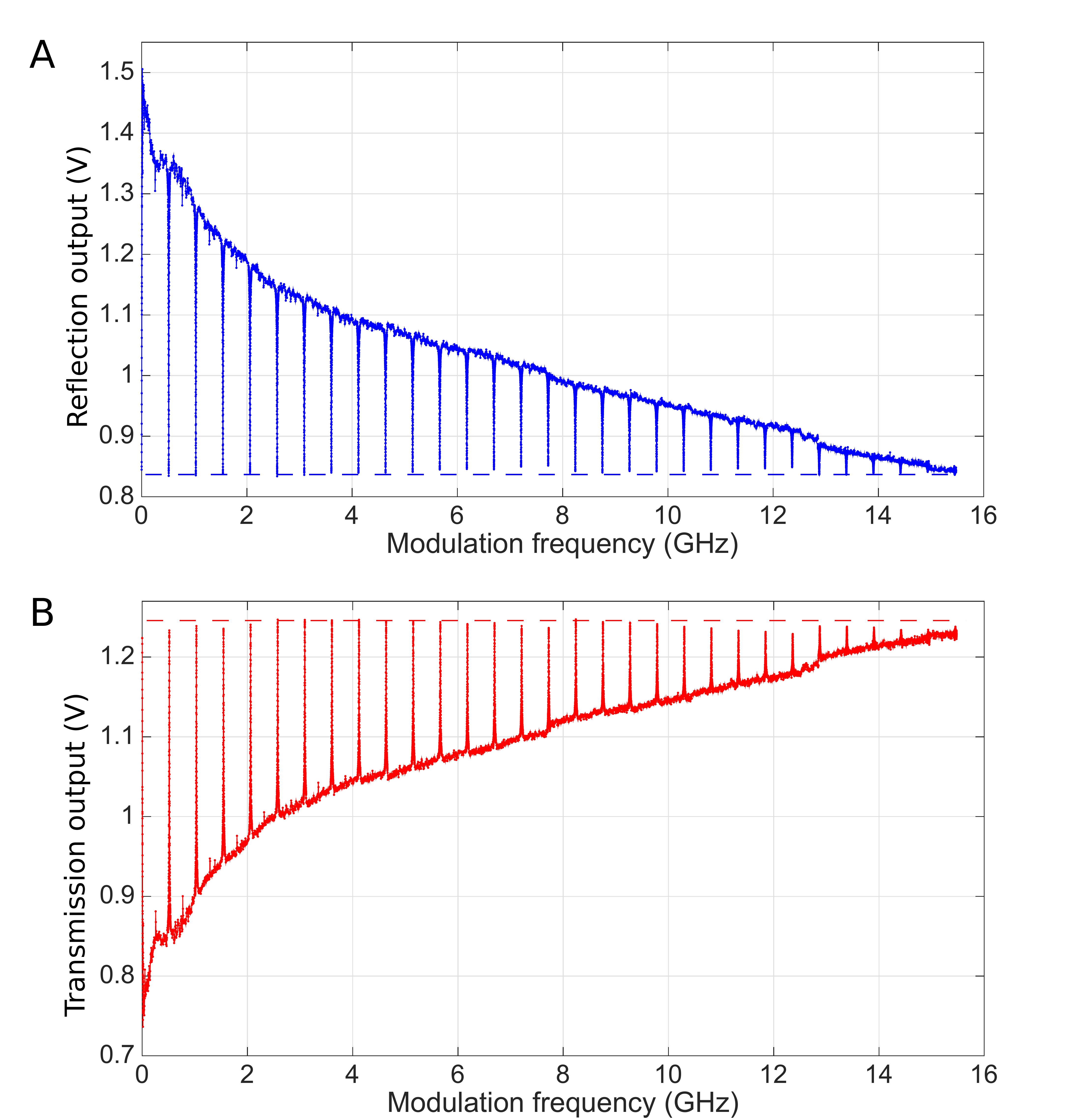}}
\caption{Frequency response of our cavity spanning 15.5$\:$GHz, showing 28 consecutive cavity resonances. A) Measured reflection output for horizontally-polarized light. B) Measured transmission output for horizontally-polarized light. The cavity was locked to the carrier frequency while the amplitude modulation frequency was swept, and the reflected/transmitted light from the cavity was captured with DC photo-detectors. The dotted lines are added to highlight the normalized on-resonance detector response, corresponding to $\mathscr{I}_{DC}^{on}$ in Eq. \ref{eqn:IdcON}.} \label{fig:ReflTx16GHz}
\end{figure}

We begin our method by capturing the cavity's transmitted and reflected light with DC photo-detectors as a function of the amplitude modulation frequency over a wide frequency range ($15.5\:$GHz). Based on the measured wide bandwidth response of the cavity shown in Fig. \ref{fig:ReflTx16GHz}, we can estimate the cavity's resonant frequencies and linewidth. This data is the normalized DC detector output, which corresponds to $\mathscr{I}_{DC}$ in Eq. \ref{eqn:IdcNorm}. The input polarization to the cavity was set to horizontally-polarized light for this wide bandwidth measurement, and the cavity was locked to the carrier frequency of this polarization. Note how the measured transmission increases to a maximum as the modulation frequency approaches the cavity's first resonance at $515\:$MHz, while the reflected intensity decreases to a minimum. The modulation frequency at the center of each peak (or dip) is equal to a multiple of the cavity FSR, while the full width at half maximum of the on-resonance data gives the linewidth. Thus from this data, we estimate the cavity's FSR to be $515\:$MHz and the linewidth to be $\sim10\:$MHz. The overall upward (downward) trend of the off-resonance data in the transmitted (reflected) data is in good agreement with the theoretical prediction shown in Fig. \ref{fig:Theoryplot}. 

Since we are using DC photo-detectors, the bandwidth of this measurement is not dependent on, or limited by, the photo-detector's response. Despite the modulator having a $\sim10\:$GHz bandwidth, we can clearly distinguish resonances at modulation frequencies up to $14.5\:$GHz. We normalize the cavity data with the modulator's output measured before the cavity to correct for any laser power drifts. We also note that the on-resonance detector levels are approximately constant, as predicted by Eq. \ref{eqn:IdcON}. In the next section, we will discuss how the modulator's response can be extracted from this cavity measurement.

\subsection{Modulator's frequency response}

\begin{figure}
\centering{\includegraphics[height=4.7 cm]{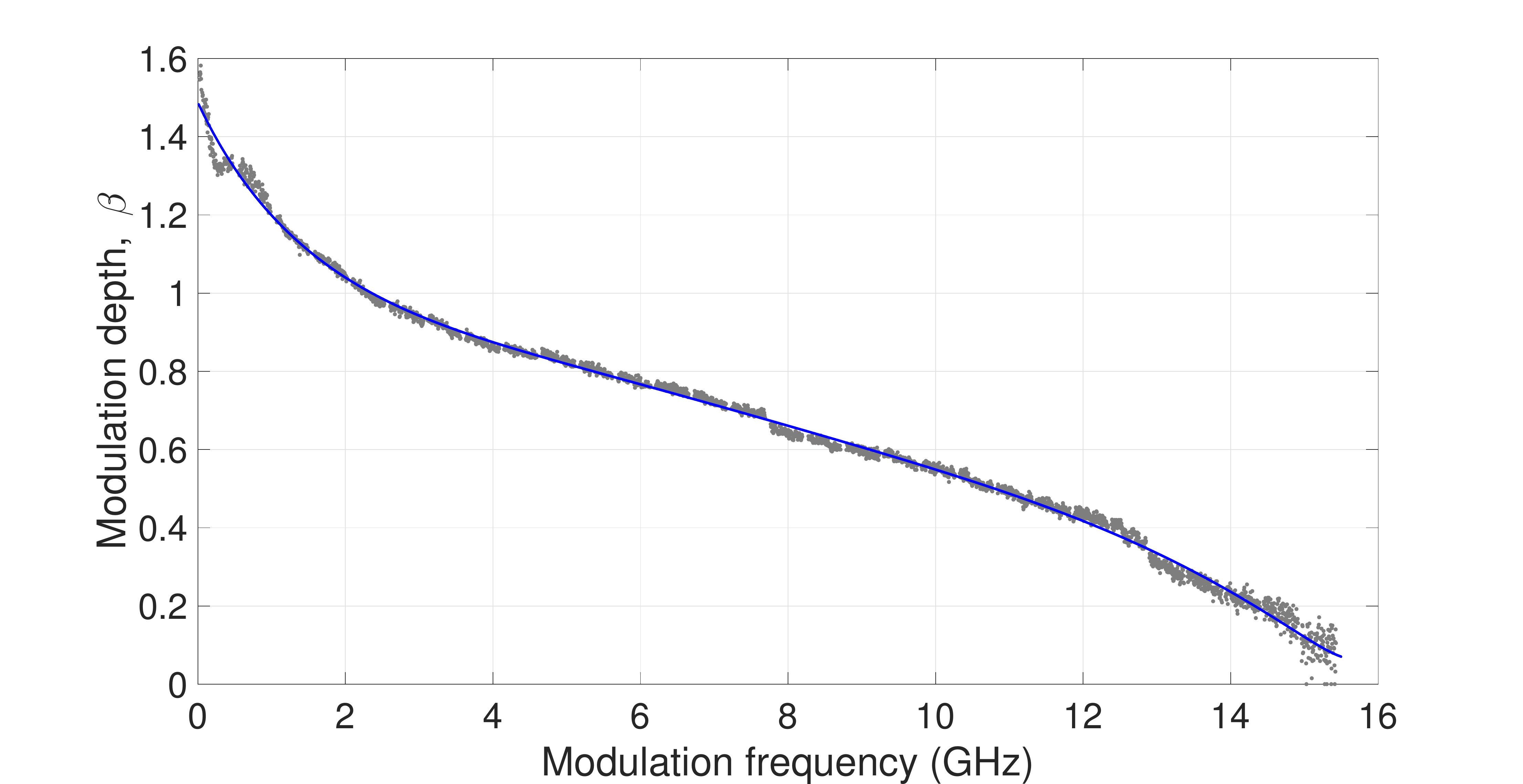}}
\caption{Frequency-dependent modulation depth estimated from the cavity's transmission response. The grey dots are experimental data, while the solid blue line is the interpolation.} \label{fig:beta}
\end{figure}

Once the cavity characteristics have been estimated, we can extract the modulator's response using the on and off-resonance data in the cavity's transmission response. The on-resonance data, corresponding to $\mathscr{I}_{DC}^{on}$, is shown as a relatively constant level in Fig. \ref{fig:ReflTx16GHz}B given by the 28 distinguishable peaks. The data inbetween these peaks corresponds to just the carrier being on resonance with the cavity while the sidebands are off resonance. Since the FSR frequency and cavity linewidth have been estimated, the on-resonance data can be removed from the transmission response to reveal just the off-resonance data, corresponding to $\mathscr{I}_{DC}^{off,T}$. We interpolate the off-resonance data with a high-order polynomial to extract the modulation depth for the entire frequency range, including at the on-resonance frequencies. 

The modulator's frequency-dependent modulation depth, $\beta$, calculated from the on and off resonance cavity data is shown in Fig. \ref{fig:beta}. Note that $\beta$ varies from 1.6 to almost 0 over the $15.5\:$GHz frequency range. A $\beta$ of 1.6 corresponds to a measurable portion of optical power present in the second harmonic modulation sidebands. Evidence of the second harmonic coupling into the cavity is visible in Fig. \ref{fig:ReflTx16GHz}A and \ref{fig:ReflTx16GHz}B as smaller resonances located halfway between the main resonances. In the next section, we will explore how the presence of second harmonic modulation sidebands in the modulator's output affects the measured cavity linewidth, and how knowing $\beta$ allows us to correct for this effect.

\subsection{Precise determination of cavity's FSR and linewidth}

\begin{figure}
\centering{\includegraphics[height=4.7 cm]{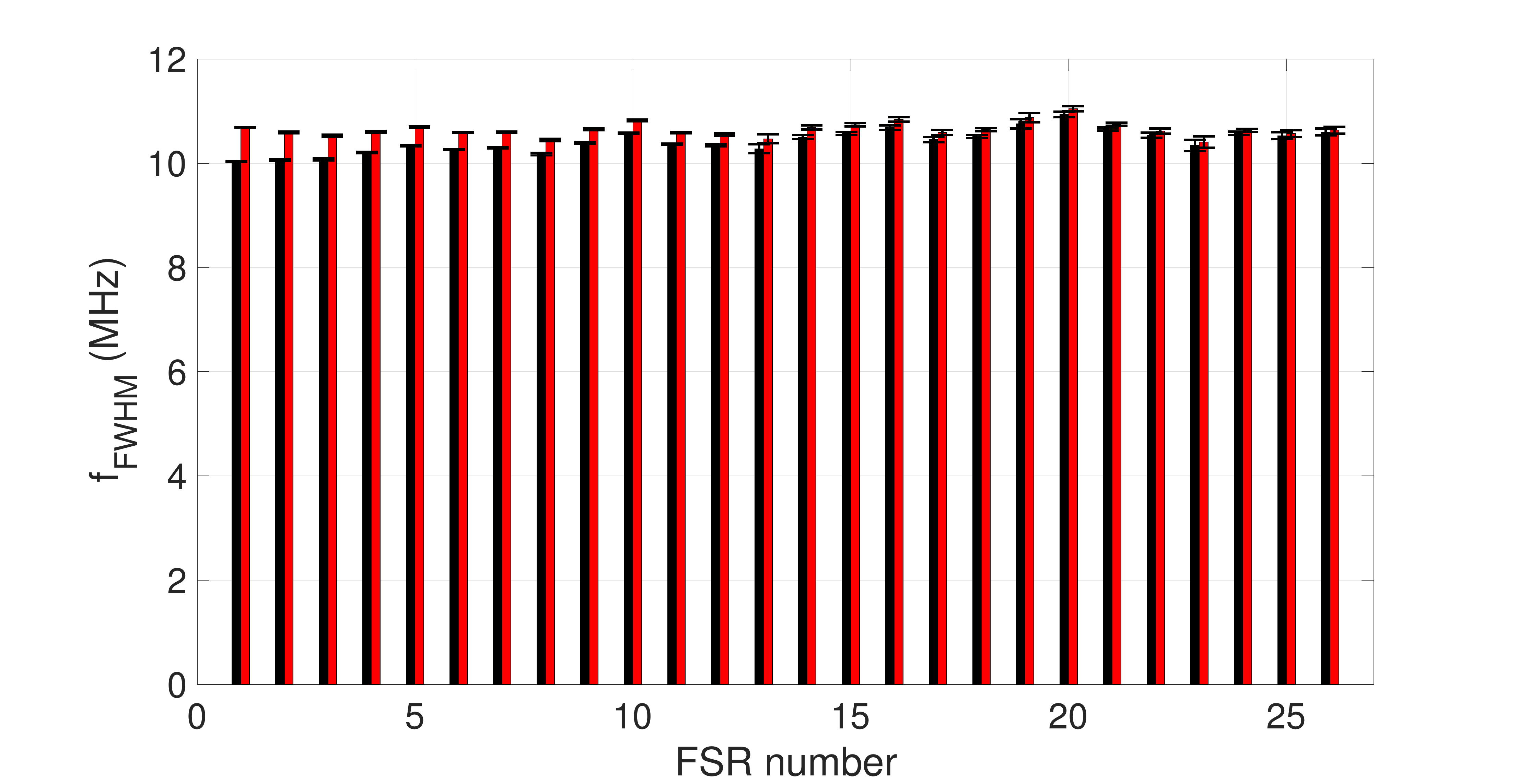}}
\caption{Linewidths (FWHM) of the first 26 consecutive cavity resonances, corresponding to the resonances shown in Fig. \ref{fig:ReflTx16GHz}. A Lorentzian function was fitted to the reflected on-resonance cavity data to more precisely determine the linewidth of each resonance. Black: fitting using a single Lorentzian function (first term in Eq. \ref{eqn:Lineshape}). Red: fitting using Eq. \ref{eqn:Lineshape}, which models both the first and second-order sidebands.} \label{fig:FWHM_refl}
\end{figure}

We use the measured modulation's response to more precisely determine the cavity's characteristics by fitting a Lorentzian function to the on-resonance data. In order to improve the accuracy of the fitting, we used an average of five data sets taken around each resonance, and normalized to the modulator's transmitted power. The amplitude modulation settings and DC bias voltage are kept constant while the frequency is swept in this narrow range. A least squares algorithm was used to fit the lineshape function (Eq. \ref{eqn:Lineshape}) to the on-resonance data to determine the FSR and linewidth. The resulting linewidths of the first 26 cavity resonances are shown in Fig. \ref{fig:FWHM_refl}, which correspond to the first 26 resonances visible in Fig. \ref{fig:ReflTx16GHz}. The linewidth results from the reflected and transmitted cavity data are quite similar so only the reflected data results are shown for clarity. 

The data in Fig. \ref{fig:FWHM_refl} shown in black are from fitting the on-resonance data with just a single Lorentzian function (first term in Eq. \ref{eqn:Lineshape}), which only models the first-order sidebands. The data shown in red corresponds to fitting with the complete lineshape function. Fitting the on-resonance data with these two methods illustrates the importance of including the second harmonic in modelling the lineshape. Since our technique measures the DC power, if modulation harmonics are present, then the detector output will consist of the sum of responses from all the sidebands. If harmonics are present but not accounted for, then the fitted linewidth will be narrower compared to the intrinsic cavity linewidth. We have quantified this effect from the frequency-dependent modulation depth, and determined the true cavity linewidth to be ($10.65\pm0.04)\:$MHz (FWHM) averaged over the 26 resonances. The error bars are calculated by individually fitting each of the five data sets, and then taking the standard deviation of the resulting five linewidths. The two fitting methods agree within error bars for frequencies above 9$\:$GHz (19th FSR) where the first-order sidebands clearly dominate. Thus, assuming the variation in $\beta$ over the cavity's linewidth is small, knowing the frequency-dependent modulator response allows us to precisely determine the characteristics of our cavity.

\begin{figure}
\centering{\includegraphics[height=4.7 cm]{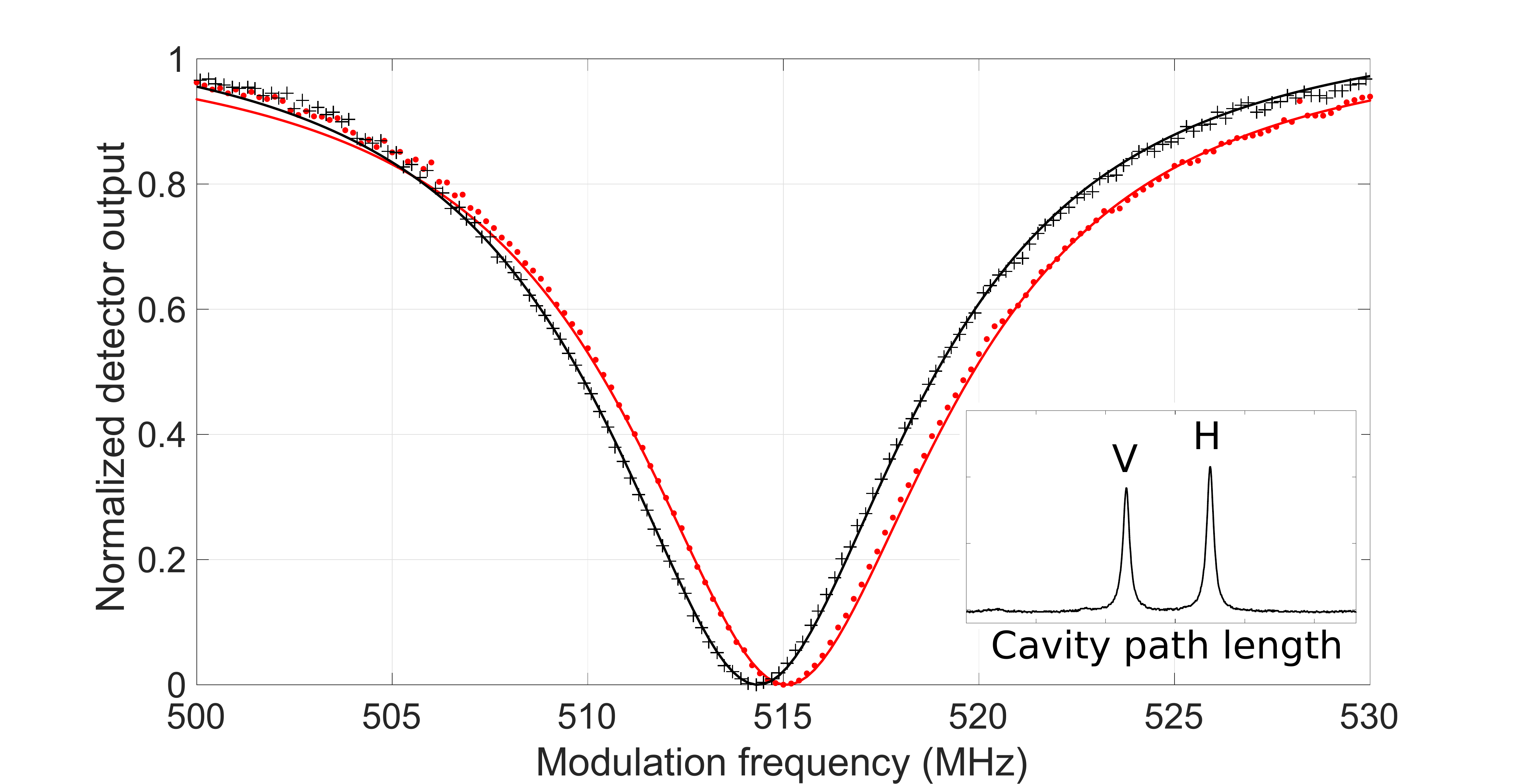}}
\caption{Frequency response of our cavity to horizontally (red dots) and vertically (black crosses) polarized light around the first resonance. The cavity was locked to the carrier frequency while the amplitude modulation frequency was swept, and the reflected light from the cavity was captured by a DC photo-detector. The data points represent measured data (normalized average of five data sets), whereas the solid lines are the Lorentzian fits (Eq. \ref{eqn:Lineshape}). The insert illustrates how the polarization modes are distinguishable and non-degenerate.} \label{fig:HVfsrs}
\end{figure}

The presence of second harmonic modulation sidebands does not corrupt the FSR measurement as all sidebands are simultaneously on resonance with the cavity. The precision and repeatability of our technique is demonstrated by our ability to measure the different resonant frequencies of orthogonal polarization modes caused by the presence of a non-linear crystal ($5\:\%$ MgO-doped LiNb$\mathrm{O}_3$) inside the cavity. The crystal has different refractive indexes along the ordinary and extra-ordinary axes, which causes horizontally-polarized light to experience a slightly different cavity path length compared to vertically-polarized light. We used a half-wave plate located before the cavity to rotate the polarization of the incident field, and characterize the distinguishable cavity modes. The two polarization modes are non-degenerate, as shown in Fig. \ref{fig:HVfsrs}, so we can lock the cavity to the carrier frequency of either polarization. Measurements of the reflected light captured around the first resonance for horizontally and vertically-polarized light are shown in Fig. \ref{fig:HVfsrs}, as well as the corresponding fit from Eq. \ref{eqn:Lineshape}. This data corresponds to the first visible resonance in the wide bandwidth measurements presented in Fig. \ref{fig:ReflTx16GHz}. We have taken the normalized average of five data sets for each polarization mode. The cavity mode due to horizontally-polarized light has a measured FSR of ($515.078\pm0.003)\:$MHz, whereas the vertically-polarized mode has a slightly lower FSR of ($514.349\pm0.004)\:$MHz. 

To illustrate the effectiveness of our measurement technique, we can use the known refractive indices of the non-linear crystal to predict the difference in resonant frequencies of the two polarization modes. The refractive indices for the ordinary and extraordinary axes of a $5\:\%$ MgO-doped LiNb$\mathrm{O}_3$ can be calculated using the temperature-independent Sellmeier equation \cite{Edwards1984},
\begin{eqnarray}
\label{eqn:no}
n_o^2 = 4.9017 + \frac{0.112280}{\lambda^2-0.049656}-0.039636\lambda^2, \\
n_e^2 = 4.5583 + \frac{0.091806}{\lambda^2-0.048086}-0.032068\lambda^2, \label{eqn:ne}
\end{eqnarray}
where $n_o$ is the refractive index of the ordinary axis, $n_e$ is the refractive index of the extraordinary axis, and $\lambda$ is the wavelength of the laser in microns (1.55032$\:\mu m$). The resonant frequency of a cavity can be calculated based on the total path length as
\begin{equation}
f_{FSR} = \frac{c}{n_1l_1+n_2l_2}, \label{eqn:FSR}
\end{equation}
where $c$ is the speed of light in a vacuum, $n_1$ is the index of refraction of air (1.000273), $l_1$ is the cavity path length in air, $n_2$ is the index of refraction of the non-linear crystal, and $l_2$ is the crystal length (10.18$\:$mm). First we can determine $l_1$ from the measured FSR for horizontally-polarized light. Then we can predict the FSR for vertically-polarized light based on this value, and the refractive index of the ordinary axis, $n_o$. Using this method, we predict an FSR of 514.384$\:$MHz, which is very close to the measured value of ($514.349\pm0.004)\:$MHz (relative difference of only $0.007\:\%$). Thus, our technique is sensitive enough to measure the effect of a path length difference of 785$\:\mu m$ ($0.13\:\%$ of the total path length) between two distinct cavity modes. The effectiveness of our measurement technique is highlighted by the fact that such an inappreciable path length difference due to the birefringent material can be precisely measured.

\section{Conclusion}

We have described a simple yet powerful calibration technique that can determine the frequency response of an optical system consisting of both an optical cavity and a high-speed amplitude modulator. We characterized both the cavity and the modulator by measuring the cavity's response over a wide frequency range with a DC photo-detector. Our method allowed us to extract the frequency-dependent modulator depth of our amplitude modulator, and characterize an optical cavity, without needing a calibrated broadband photo-detector or optical spectrum analyzer. We used the on and off-resonance cavity data to precisely identify the intrinsic cavity linewidth, which would otherwise be corrupted by the presence of higher-order modulation harmonics from the amplitude modulator. In addition, we demonstrated the precision and repeatability of our technique by measuring the different resonant frequencies of orthogonal polarization cavity modes. Once the modulator has been characterized, our method can be applied to characterize any passive optical element including, by not limited to, cavities.

\section*{Funding}
This work was supported financially by the Australian Research Council Centres of Excellence scheme number CE110001027, the Office of Naval Research (ONR), and Industry Canada.

\section*{Acknowledgments}
The authors would like to thank Greg Milford for lending us the RF signal generator, and Darryl Budarick, Shane Brandon, and Mitchell Sinclair for much appreciated technical support. We would also like to thank David Moilanen for fruitful discussions.



\end{document}